%% file: paper.tex
\documentclass[conference]{IEEEtran}

\usepackage{cite}
\usepackage{hyperref}
\usepackage[utf8]{inputenc}
\usepackage[frozencache]{minted}
\usepackage[dvipsnames]{xcolor}
\usepackage{amsthm}
\usepackage{graphicx}
\usepackage{caption}
\usepackage{subcaption}

\setminted{fontsize=\footnotesize}

\theoremstyle{definition}
\newtheorem{example}{Example}
\newtheorem{snippet}{Snippet}

\definecolor{rosy}{HTML}{ff0066}
\definecolor{ros}{HTML}{6600cc}

\newcommand{\rosy}{{\color{rosy}\textsc{rosy}}}
\newcommand{\roshask}{{\color{ros}\textsc{roshask}}}
\newcommand{\ROS}{{\color{ros}\textsc{ROS}}}

\begin{document}

\title{\rosy: An elegant language to teach the pure reactive nature of robot programming}

\author{\IEEEauthorblockN{Hugo Pacheco}
\IEEEauthorblockA{Universidade do Minho \& INESC TEC\\
Portugal\\
Email: hpacheco@di.uminho.pt}
\and
\IEEEauthorblockN{Nuno Macedo}
\IEEEauthorblockA{Universidade do Minho \& INESC TEC\\
Portugal\\
Email: nfmmacedo@di.uminho.pt}
}

\maketitle

\begin{abstract}

Robotics is incredibly fun and is long recognized as a great way to teach programming, while drawing inspiring connections to other branches of engineering and science such as maths, physics or electronics.
Although this symbiotic relationship between robotics and programming is perceived as largely beneficial, educational approaches often feel the need to hide the underlying complexity of the robotic system, but as a result fail to transmit the reactive essence of robot programming to the roboticists and programmers of the future.

This paper presents \rosy, a novel language for teaching novice programmers through robotics.
Its functional style is both familiar with a high-school algebra background and a materialization of the inherent reactive nature of robotic programming.
Working at a higher-level of abstraction also teaches valuable design principles of decomposition of robotics software into collections of interacting controllers.
Despite its simplicity, \rosy\ is completely valid Haskell code compatible with the \ROS~ecosystem.

We make a convincing case for our language by demonstrating how non-trivial applications can be expressed with ease and clarity, exposing its sound functional programming foundations, and developing a web-enabled robot programming environment.

\end{abstract}

\IEEEpeerreviewmaketitle

\input{intro}
\input{relwork}

\input{lang}
\input{env}
\input{impl}

\section{Conclusion} \label{sec:conc}

In this paper we have presented \rosy, a new pedagogical robot programming language that advocates a sweet-spot between the expressiveness of FRP and the needed simplicity of an educational setting.
As part of a computing summer camp for held at the University of Minho\footnote{\url{https://www.uminho.pt/veraonocampus}}, in July 2019, we have taught a 4-hour training session for K12 students on hands-on robot programming in \rosy, with only two prior sessions of general programming in Haskell.
From our perceptions, students were able to comprehend the concepts and quickly start programming the robot to perform simple tasks.

In the future, we plan to provide further \rosy~training sessions and undergo empirical studies that can corroborate its practical value for learning programming via robotics.
We plan to improve the \rosy~environment with more advanced simulation scenarios using, e.g., the Gazebo web client\footnote{http://gazebosim.org/gzweb.html} or remote \ROS~support similar to~\cite{CasanCMAM2015}.
We also plan to explore the design of novel novice-friendly interfaces that blend textual and visual representations, and graphically combine visual blocks typical of imperative robotic approaches with graphical data flow diagrams typical of advanced FRP approaches.


\bibliographystyle{IEEEtran}
\bibliography{IEEEabrv,paper}

\end{document}

%% file: intro.tex
\section{Introduction}

Robotics, though a multi-disciplinary area of engineering and science, shares a unique symbiotic relationship with computer science.
As robots are increasingly put to solve more complex tasks, the more important it becomes to look carefully into the software that runs inside them, and in particular to the programming languages that bring them to life.
At the same time, programming a robot can grant  computer science an often missing practical appeal, with a pedagogical potential long recognized by computer science educators~\cite{Blank2006}.

Various studies corroborate that robotics can indeed  mitigate the abstract nature of programming~\cite{oddie2010introductory} and provide a more creative environment for teaching programming concepts focused on hands-on problem solving~\cite{druin2000robots} and for coding rich behavior using basic code structures~\cite{gandy2010use}, while still requiring reasoning about relevant computational concepts such as software modularity and communication~\cite{LawheadDBGSBH:03}.

Yet, the richness of robot programming is synonym of its real-world complexity, which renders standard robotic frameworks unfeasible for use in a pedagogical setting and has been motivating the proposal of various educational robot programming frameworks.
As the synergy between robotics and programming deepens, education is the perfect place to experiment novel approaches that can shape the robot programming languages of the future~\cite{cleary2015reactive}.
It also carries a timely opportunity to transmit good design practices to new generations of programmers, that are receptive to novel languages as long as these allow to quickly build applications~\cite{felleisen2004teachscheme}.

As a step towards realising this vision, we advocate that languages for teaching robotics to novice programmers shall:
\begin{itemize}
    \item be compatible with standard robotic practices and seamlessly connect to existing robotic infrastructures;
    \item adopt a simple declarative programming style and provide a pure cause-and-effect interface emphasizing the essence of what it means to program a reactive system;
    \item rely on a general-purpose programming language with good tool support, so that advanced programming features can be gradually introduced and acquired programming skills can naturally transfer to other domains.
\end{itemize}
In Section~\ref{sec:rel}, we argue that state-of-the-art languages fail, to some extent, to exhibit these characteristics. This justifies the proposal of \rosy, a simple yet powerful reactive programming language, presented in Section~\ref{sec:lang}. To ease adoption by novice programmers, \rosy~is supported by a browser-based development environment, described in Section~\ref{sec:ide}. As an embedded domain-specific language, \rosy~supports the full power of higher-order functional programming offered by the host Haskell language, and is connected to \ROS, one of the most popular robotic middlewares. The mechanics behind its implementation is presented in Section~\ref{sec:tech}. Section~\ref{sec:conc} concludes the paper and leaves directions for future work.

%% file: relwork.tex
\section{The pedagogy of robot programming languages} \label{sec:rel}

Programming robots is a particularly complex task, where one has to deal not only with the continuous and real-time aspects of the physical world, but also with heterogeneous architectures and complex communication paradigms.
To minimize frustration of novice programmers, several pedagogical languages and approaches have been proposed over the years.
This section reviews and discusses such related work.

\subsection{The Robot Operating System}
Robotic middlewares have been developed to ease the programming of robots, abstracting hardware and communication details and promoting modularity. The Robot Operating System (\ROS)~\cite{QuigleyCGFFLWN:09} is one such middleware, possibly the most popular, and defines an architecture through which components, called \emph{nodes} in \ROS, can communicate with each other by publishing to or subscribing from data sources, called \emph{topics}. Other than that, individual nodes are programmed in general-purpose languages, typically C++ or Python. The popularity of \ROS~is fueled by a very dynamic community and an open-source policy that encourages code re-use, and a large package database ranging from educational to industrial applications.

As a pedagogical example, consider the well-known TurtleBot2 robot, whose controller is programmed in \ROS-powered C++. A main node controls its Kobuki mobile base publishing odometry information and data collected from a set of sensors (collision, cliff and wheel drop sensors, plus a few buttons), and subscribing to commands that control its (linear and angular) velocity and the color of a set of LEDs. \ROS~allows the definition of custom message types, from which source code is automatically generated, and Kobuki defines such message types for the relevant events. Below is a minimal example of a \ROS~application that subscribes to bumper events and plays an error sound when a bumper is pressed:
\begin{snippet}[Play a sound on collision]~
\begin{minted}[escapeinside=@@]{cpp}
@\textbf{\color{ros}{ros::Publisher}}@ pub;

void cb (const @\textbf{\color{ros}{kobuki\_msgs::BumperEventConstPtr}}@& b){
  if (b->@\color{ros}{state}@==@\textbf{\color{ros}{kobuki\_msgs::BumperEvent::PRESSED}}@){
    @\textbf{\color{ros}{kobuki\_msgs::Sound}}@ s;
    s.@\color{ros}{value}@ = @\textbf{\color{ros}{kobuki\_msgs::Sound::ERROR}}@;
    pub.@\color{ros}{publish}@(s); } }
  
int main(int argc, char** argv){
  @\color{ros}{ros::init}@(argc, argv, "play");
  @\textbf{\color{ros}{ros::NodeHandle}}@ nh;
  @\textbf{\color{ros}{ros::Subscriber}}@ sub = nh.@\color{ros}{subscribe}@(
           "/mobile_base/events/bumper", 10, cb);
  pub = nh.@\color{ros}{advertise}@<@\textbf{\color{ros}{kobuki\_msgs::Sound}}@>(
           "/mobile_base/commands/sound", 10);
  @\color{ros}{ros::spin}@();
  return 0; }
\end{minted}
\label{snip:play}
\end{snippet}
\noindent
Even this minimal snippet is clearly non-trivial for novice programmers, obfuscating the truly reactive nature of the controller.
Besides having to get around advanced linguistic features such as pointers, namespaces or templates, to manipulate topics the programmer must first explicitly \mintinline[escapeinside=@@]{cpp}{@\color{ros}{subscribe}@} to bumper events and \mintinline[escapeinside=@@]{cpp}{@\color{ros}{advertise}@} that sound commands will be published, considering the buffer size for incoming and outgoing messages.
The programmer must also reason about how topics are processed, by registering a callback on the \mintinline[escapeinside=@@]{cpp}{@\textbf{\color{ros}{ros::Subscriber}}@} that will \mintinline[escapeinside=@@]{cpp}{@\color{ros}{publish}@} an error sound command if a bumper pressed event is read, and when topics are processed, in this simplest case using the \ROS~\mintinline[escapeinside=@@]{cpp}{@\color{ros}{spin}@} primitive that periodically processes callbacks for the queued messages.

\subsection{Visual robot programming languages}

\begin{figure}[t]
\centering
    \includegraphics[width=0.6\columnwidth]{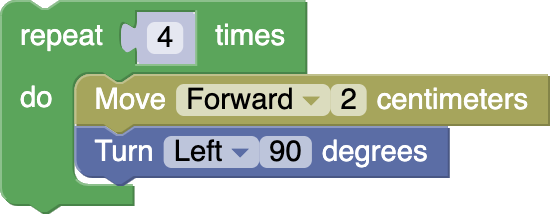}
  \caption{A simple block that draws a square.}
  \label{fig:block}
\end{figure}

To tame the complexity of programming robots in fully-fledged general-purpose programming languages, a myriad of frameworks aimed at novice robot programming using visual programming languages have been proposed~\cite{DiproseMH2011}. Many of these follow a block-based approach~\cite{CasanCMAM2015,angulo2017roboblock,masum2018framework,WeintropASFLSF:18,marghitu2015robotics}, where predefined blocks with varied colors and edges can be put together like pieces of a puzzle to define a robotic controller.

Aside from discussions on whether block-based approaches constitute ``real'' programming or their programming experience transfers to ``real'' textual languages~\cite{weintrop2019block,xu2019block}, and with due exceptions such as~\cite{marghitu2015robotics}, the vast majority of block-based robot programming approaches adopt an imperative mindset: primitive blocks execute individual predefined tasks, such as moving forward during a certain period of time or for a certain distance; and combining blocks amounts to performing sequences of tasks.
For example, the TurtleBot3 can be programmed via a block-based interface in which we can write a block similar to the one from Figure~\ref{fig:block}. Albeit simple, this example block hides away the reactive nature of the robotic system -- all the sensor and command behavior is encapsulated within the black-box primitive blocks.

\subsection{Functional reactive programming languages}

Reactive programming~\cite{bainomugisha2013survey} is a programming paradigm organised around information producers and consumers, that can naturally bring out the intrinsically reactive nature of cyber-physical systems.
A particularly active line of research, known as functional reactive programming (FRP)~\cite{perez2018functional}, focuses on streams of information as the central reactive abstraction, and advocates a declarative approach to manipulate streams at a high-level of abstraction supported by the pure equational reasoning of functional languages such as Haskell.

Functional languages, as a form of algebra, are a good fit for introductory programming~\cite{felleisen2004teachscheme}, and many pedagogical FRP approaches have been proposed for programming interactive games and animations~\cite{felleisen2009functional,codeworld,almeida2018teaching,cleary2015reactive}.
Much classical work has also proposed FRP for programming robots~\cite{peterson1999lambda,hudak2002arrows,pembeci2002functional}.

In FRP, time is typically explicit and conceptually continuous; the execution of the system is then carried out by sampling all streams synchronously at the rate of an external and global clock.
The \roshask~\cite{roshask} Haskell \ROS~library promotes the modularity and expressiveness of FRP, while remaining faithful to the asynchronous nature of \ROS, by adopting a more pragmatic approach centered on manipulating asynchronous topics in their entirety.

\subsection{Towards a pedagogical robot FRP language}

Despite the declarative nature of functional programming and the focus on events of reactive programming, the combinatorial FRP style is not beginner friendly, as it tends to swallow entire programs and resorts to advanced higher-order features to separate reactive code (referring to entire topics) from non-reactive code (referring to individual events).

In this paper, we advocate that a toned-down FRP language, focused on individual events, represents a sweet spot of introductory robot programming.
Making time implicit is a big win, as it liberates novice programmers from specific FRP syntax and invites them to simply write pure functions.
To retain some of the expressiveness of FRP, these functions have then an intuitive interpretation as operations on data streams.

%% file: lang.tex
\section{The \rosy~language} \label{sec:lang}

The \rosy~language presents itself as a natural dialect for bringing robots alive using nothing more than plain mathematics, while promoting good software design practices.
In this section, we make a case for how its declarative nature allows creating robotic controllers in an intuitive and painless way, and informally present its defining features through a collection of examples of increasing complexity that illustrate how to control a TurtleBot2 robot. The types and associated fields used in the examples will therefore spell the standard Kobuki \ROS~message types.
The \rosy~website\footnote{\url{https://haskell.lsd.di.uminho.pt/rosy}} offers a modern integrated development environment, including an editor, help guides, and executable versions of all the examples shown in this section and more.

For readers not familiar with Haskell syntax, all the functions and operators not defined in this paper are standard and their definition can be found at \url{https://hoogle.haskell.org}.
The documentation for respectively colored \rosy-specific functions and types is available at the \rosy~website.

\subsection{A \rosy~primer}

In \rosy, we can control a robot by writing pure functions that receive sensor information from the robot and react by sending commands back to the robot.
To make a robot move forward at a constant velocity of 0.5 m/s, we can simply write:
\begin{example}[Move forward]~
\begin{minted}[escapeinside=@@]{haskell}
move :: @\textbf{\color{rosy}{Velocity}}@
move = @\textbf{\color{rosy}{Velocity}}@ 0.5 0

main = @\color{rosy}{simulate}@ move
\end{minted}
\end{example}
\noindent
The\mintinline{haskell}{ move} function models our controller (where the \mintinline[escapeinside=@@]{haskell}{@\textbf{\color{rosy}{Velocity}}@} of the robot is separated into its linear and angular components), and the\mintinline{haskell}{ main} function is \rosy-specific syntax to\mintinline[escapeinside=@@]{haskell}{ @\color{rosy}{simulate}@} our controller (that will be elided from now on).
Still, the astute reader may fittingly ask ``for how long are we telling the robot to move forward?''
Being \rosy~a reactive programming language, the answer is not ``once'' or ``for a certain period of time'', but actually ``forever''.
The intuition is that a controller is a function that unceasingly listens for inputs, and for each received input produces an output, what grants \rosy~programs an implicit notion of time.
In this case, since\mintinline{haskell}{ move} receives no inputs, it will produce \mintinline[escapeinside=@@]{haskell}{@\textbf{\color{rosy}{Velocity}}@} outputs at a fixed rate.

To make things a bit more interesting, imagine that we want the robot to accelerate forward with non-constant velocity. We can achieve this behavior by making sure to increase the robot's velocity at each point in time:
\begin{example}[Accelerate forward]~
\begin{minted}[escapeinside=@@]{haskell}
accelerate :: @\textbf{\color{rosy}{Velocity}}@ -> @\textbf{\color{rosy}{Velocity}}@
accelerate (@\textbf{\color{rosy}{Velocity}}@ vl va) = @\textbf{\color{rosy}{Velocity}}@ (vl+0.5) va
\end{minted}
\end{example}
\noindent
Note that the same \mintinline[escapeinside=@@]{haskell}{@\textbf{\color{rosy}{Velocity}}@} type has different input and output meanings.
This second\mintinline{haskell}{ accelerate} controller is a function that repeatedly asks the robot for its current linear velocity, and commands the robot to increase it by 0.5 m/s.

As our robot is moving forward, what if it hits a wall?
We can naturally express multiple \rosy~controllers that react to distinct events.
For instance, we can make the robot play an error sound when one of its bumpers is pressed, what will happen on contact with a wall:
\begin{example}[Accelerate and play a sound on collision]~
\begin{minted}[escapeinside=@@]{haskell}
play :: @\textbf{\color{rosy}{Bumper}}@ -> Maybe @\textbf{\color{rosy}{Sound}}@
play (@\textbf{\color{rosy}{Bumper}}@ _ @\textbf{\color{rosy}{Pressed}}@)  = Just @\textbf{\color{rosy}{ErrorSound}}@
play (@\textbf{\color{rosy}{Bumper}}@ _ @\textbf{\color{rosy}{Released}}@) = Nothing

accelerateAndPlay = (accelerate,play)
\end{minted}
\label{ex:3}
\end{example}
\noindent
In this example, we define a composite\mintinline{haskell}{ accelerateAndPlay} controller by simply pairing together\mintinline{haskell}{ accelerate} and\mintinline{haskell}{ play}.
Note that these two functions are not required to execute at the same time:\mintinline{haskell}{ accelerate} runs on periodic robot information, though\mintinline{haskell}{ play} waits for \mintinline[escapeinside=@@]{haskell}{@\textbf{\color{rosy}{Bumper}}@} events.
The two controllers will execute in parallel, effectively combining both behaviors.
To be able to play a sound only when a bumper is \mintinline[escapeinside=@@]{haskell}{@\textbf{\color{rosy}{Pressed}}@}, and not \mintinline[escapeinside=@@]{haskell}{@\textbf{\color{rosy}{Released}}@}, the\mintinline{haskell}{ play} function may or may not produce a \mintinline[escapeinside=@@]{haskell}{@\textbf{\color{rosy}{Sound}}@} command. The\mintinline{haskell}{ play} \rosy~controller displays the same behavior as the \ROS~one encoded in Snippet~\ref{snip:play}, but here its reactive nature is clear from the type declaration.

Even though the controller from Example~\ref{ex:3} detects when the robot hits a wall, it will continue to push against the wall, likely reaching a deadlock.
With controllers as pure functions that react to events as they happen, there is no easy way to make a controller remember some event in the past.
This contrasts with traditional imperative robot programming languages, where we could use a global variable to, e.g., memorize when the robot has hit a wall and, from then on, change its behavior.
Like other pedagogical functional languages~\cite{felleisen2009functional}, we grant \rosy~controllers a notion of \mintinline[escapeinside=@@]{haskell}{@\textbf{\color{rosy}{Memory}}@}, that can be used to, e.g., remember the robot's moving direction:
\begin{example}[Accelerate forward and backwards on collision]~
\begin{minted}[escapeinside=@@]{haskell}
type Hit = Bool

reverseDir :: @\textbf{\color{rosy}{Bumper}}@ -> @\textbf{\color{rosy}{Memory}}@ Hit
reverseDir _ = @\textbf{\color{rosy}{Memory}}@ True

accelerate :: @\textbf{\color{rosy}{Memory}}@ Hit -> @\textbf{\color{rosy}{Velocity}}@ -> @\textbf{\color{rosy}{Velocity}}@
accelerate (@\textbf{\color{rosy}{Memory}}@ hit) (@\textbf{\color{rosy}{Velocity}}@ vl va) = if hit
    then @\textbf{\color{rosy}{Velocity}}@ (vl-0.5) va
    else @\textbf{\color{rosy}{Velocity}}@ (vl+0.5) va
    
forwardBackward = (reverseDir,accelerate)
\end{minted}
\end{example}
\noindent
In the above example, the controller hands over its memory to\mintinline{haskell}{ accelerate}, that uses it to determine in which direction to move. To reconcile memory with pure functional programming, controllers that wish to change the memory are expected to return it explicitly as an output. The \mintinline{haskell}{Hit} boolean memory will be false by default, and changed to true by\mintinline{haskell}{ reverseDir} when a bumper event occurs.

\begin{figure}[t]
    \centering
    \includegraphics[width=\columnwidth]{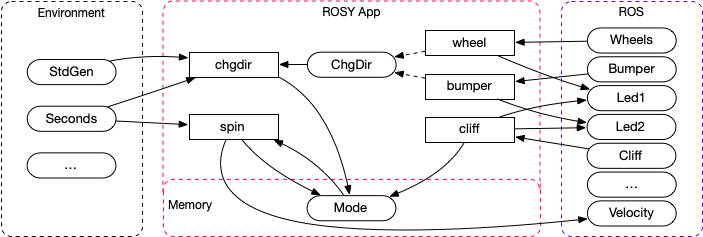}
    \caption{The random walker application in \rosy.}
    \label{fig:rosy_app}
\end{figure}

\subsection{Revisiting \ROS~controllers}

For a more complete and realistic example, we now encode the popular Kobuki random walker controller in \rosy, while trying to stay faithful to the C++ \ROS~implementation\footnote{\url{https://github.com/yujinrobot/kobuki/tree/devel/kobuki_random_walker}}:
\begin{itemize}
\item On a bumper or cliff event, the robot blinks one of its LEDs orange and decides to change its direction;
\item On a wheel drop event, the robot blinks both LEDs red and decides to stop moving while the wheel is in the air;
\item When changing direction, it randomly decides on an angle between $0^\circ$ and $180^\circ$ and on a left/right direction. Depending on a fixed angular velocity, it estimates how many seconds it shall turn;
\item A sequential loop routinely performs the adequate action depending on the state of the controller. It commands the robot to either: go forward, stop moving, or turn in a given direction for a number of seconds.
\end{itemize}
\begin{example}[Random walker]~
\begin{minted}[escapeinside=;;]{haskell}
data Mode = Go | Stop | Turn Double ;\textbf{\color{rosy}{Seconds}};
data ChgDir = ChgDir -- change direction

vel_lin = 0.5
vel_ang = 0.1

bumper :: ;\textbf{\color{rosy}{Bumper}}; -> (;\textbf{\color{rosy}{Led1}};,Maybe ChgDir)
bumper (;\textbf{\color{rosy}{Bumper}}; _ st) = case st of
  ;\textbf{\color{rosy}{Pressed}}; -> (;\textbf{\color{rosy}{Led1}}; ;\textbf{\color{rosy}{Orange}};,Just ChgDir)
  ;\textbf{\color{rosy}{Released}}; -> (;\textbf{\color{rosy}{Led1}}; ;\textbf{\color{rosy}{Black}};,Nothing)

cliff :: ;\textbf{\color{rosy}{Cliff}}; -> (;\textbf{\color{rosy}{Led2}};,Maybe ChgDir)
cliff (;\textbf{\color{rosy}{Cliff}}; _ st) = case st of
  ;\textbf{\color{rosy}{Hole}}; -> (;\textbf{\color{rosy}{Led2 Orange}};,Just ChgDir)
  ;\textbf{\color{rosy}{Floor}}; -> (;\textbf{\color{rosy}{Led2 Black}};,Nothing)

wheel :: ;\textbf{\color{rosy}{Wheel}}; -> (;\textbf{\color{rosy}{Led1}};,;\textbf{\color{rosy}{Led2}};,;\textbf{\color{rosy}{Memory}}; Mode)
wheel (;\textbf{\color{rosy}{Wheel}}; _ st) = case st of
  ;\textbf{\color{rosy}{Air}}; -> (;\textbf{\color{rosy}{Led1}}; ;\textbf{\color{rosy}{Red}};,;\textbf{\color{rosy}{Led2}}; ;\textbf{\color{rosy}{Red}};,;\textbf{\color{rosy}{Memory}}; Stop)
  ;\textbf{\color{rosy}{Ground}}; -> (;\textbf{\color{rosy}{Led1}}; ;\textbf{\color{rosy}{Black}};,;\textbf{\color{rosy}{Led2}}; ;\textbf{\color{rosy}{Black}};,;\textbf{\color{rosy}{Memory}}; Go)

chgdir :: ChgDir -> StdGen -> ;\textbf{\color{rosy}{Seconds}};
       -> ;\textbf{\color{rosy}{Memory}}; Mode
chgdir _ r now = ;\textbf{\color{rosy}{Memory}}; (Turn dir time)
    where
    (b,r') = random r
    (ang,_) = randomR (0,pi) r'
    dir = if b then 1 else -1
    time = now + ;\color{rosy}{doubleToSeconds}; (ang/vel_ang)

spin :: ;\textbf{\color{rosy}{Memory}}; Mode -> ;\textbf{\color{rosy}{Seconds}};
     -> (;\textbf{\color{rosy}{Velocity}};,;\textbf{\color{rosy}{Memory}}; Mode)
spin m@(;\textbf{\color{rosy}{Memory}}; Stop) _ = (Velocity 0 0,m)
spin m@(;\textbf{\color{rosy}{Memory}}; (Turn dir t)) now | t > now =
  (;\textbf{\color{rosy}{Velocity}}; 0 (dir*vel_ang),m)
spin m _ = (Velocity vel_lin 0,;\textbf{\color{rosy}{Memory}}; Go)

randomWalk = (bumper,cliff,wheel,chgdir,spin)
\end{minted}
\end{example}
\noindent
The\mintinline{haskell}{ randomWalk} controller encodes each part of the above specification as a separate function, sharing a memory \mintinline{haskell}{Mode} that encodes the different robot states. The \ROS-style computation graph is illustrated in Fig.~\ref{fig:rosy_app}, where nodes amount to the defined reactive functions, and topics are either user-defined events or those provided for the Kobuki controller.
The\mintinline{haskell}{ bumper} and\mintinline{haskell}{ cliff} functions change the LED colors, and set in motion a change in direction. For greater modularity, they both emit a new event of type \mintinline{haskell}{ChgDir}.
The\mintinline{haskell}{ wheel} function also changes the LEDs' colors and sets the memory mode to the \mintinline{haskell}{Stop} state.
The\mintinline{haskell}{ spin} function reads the memory mode, sets the respective velocity depending on the mode, and returns an updated mode. It receives the current time in \mintinline[escapeinside=@@]{haskell}{@\textbf{\color{rosy}{Seconds}}@} to determine if the estimated turning time has elapsed.

The most complicated behavior is left to function\mintinline{haskell}{ chgdir}, that resolves \mintinline{haskell}{ChgDir} events to concrete \mintinline{haskell}{Turn} actions.
To implement random behavior, it resorts to a Haskell standard randomness generator of type \mintinline{haskell}{StdGen} to generate a random direction and angle, and reads the current time to calculate the time limit for the \mintinline{haskell}{Turn} action.
This is a good example of how the power of the full Haskell language can be gradually unleashed as students tackle more advanced problems.

Another popular Kobuki controller is the safety controller\footnote{\url{https://github.com/yujinrobot/kobuki/tree/devel/kobuki_safety_controller}}, that imposes stricter conditions on dangerous events.
Its C++ \ROS~implementation can be encoded in \rosy~as a single controller that reacts to bumper, cliff or wheel drop events and cautiously decides on a new velocity to escape danger:
\begin{example}[Safety controller]~
\begin{minted}[escapeinside=@@]{haskell}
safetyControl :: Either (Either @\textbf{\color{rosy}{Bumper}}@ @\textbf{\color{rosy}{Cliff}}@) @\textbf{\color{rosy}{Wheel}}@
     -> Maybe @\textbf{\color{rosy}{Velocity}}@
safetyControl = ...
\end{minted}
\end{example}
\noindent
The code for \mintinline{haskell}{ safetyControl} is conceptually simple, yet verbose as it explores multiple combinations of sensor inputs. We omit it in the paper, but it can be found at the \rosy~website.

The safety controller does not do much by itself, and it is typically deployed together with the random walker to limit its actions as the robot roams around.
Since both controllers publish possibly conflicting \mintinline[escapeinside=@@]{haskell}{@\textbf{\color{rosy}{Velocity}}@} commands to the robot, the traditional \ROS~solution is to use a multiplexer that remaps topics and allows one controller at a time to command the robot, according to a fixed set of priorities.

We can define a general multiplexer in \rosy~as follows:
\begin{example}[Binary multiplexer, \mintinline{haskell}{M1} with priority over \mintinline{haskell}{M2}]~
\begin{minted}[escapeinside=@@]{haskell}
data M = Start | Ignore @\textbf{\color{rosy}{Seconds}}@
data M1 a = M1 a
data M2 a = M2 a

mux :: @\textbf{\color{rosy}{Seconds}}@ -> @\textbf{\color{rosy}{Memory}}@ M
    -> Either (M1 a) (M2 a) -> Maybe (a,@\textbf{\color{rosy}{Memory}}@ M)
mux t _ (Left (M1 a)) = Just (a,@\textbf{\color{rosy}{Memory}}@ (Ignore(t+d)))
mux t (@\textbf{\color{rosy}{Memory}}@ (Ignore s)) _ | s > t = Nothing
mux t _ (Right (M2 a)) = Just (a,@\textbf{\color{rosy}{Memory}}@ Start)
\end{minted}
\end{example}
\noindent
This binary multiplexer reads the current time\mintinline{haskell}{ t} in \mintinline[escapeinside=@@]{haskell}{@\textbf{\color{rosy}{Seconds}}@} and reacts to events either marked as \mintinline{haskell}{M1} or \mintinline{haskell}{M2}, giving higher priority to \mintinline{haskell}{M1} events by setting a time interval starting at\mintinline{haskell}{ t} and ending at\mintinline{haskell}{ t+d} (for a fixed duration\mintinline{haskell}{ d}) during which all \mintinline{haskell}{M2} events are ignored.

We can then instantiate a safe random walker by remapping output velocities of the safety and random walker controllers with \mintinline{haskell}{M1} and \mintinline{haskell}{M2} tags and running a multiplexer in parallel:
\begin{example}[Random walker with safety controller]~
\begin{minted}[escapeinside=@@]{haskell}
safetyControl :: ... -> Maybe (M1 @\textbf{\color{rosy}{Velocity}}@)
spin :: ... -> (M2 @\textbf{\color{rosy}{Velocity}}@,...)
muxVel :: ... -> Either (M1 @\textbf{\color{rosy}{Velocity}}@) (M2 @\textbf{\color{rosy}{Velocity}}@)
       -> Maybe (@\textbf{\color{rosy}{Velocity}}@,...)
    
safeRandomWalk = (randomWalk,safetyControl,muxVel)
\end{minted}
\end{example}
\noindent
The complete refactored code for the\mintinline{haskell}{ safeRandomWalk} controller is available at the \rosy~website.

\subsection{Revisiting block-based languages}

Because controllers in \rosy~run forever, they do not directly lend themselves to performing sequences of instructions.
For concreteness, imagine that we want to command the robot to draw a square on the floor with its movement.
In a visual robot programming language, this can be done by assembling a block like the one from Fig.~\ref{fig:block}.
Even though such a block can be expressed in \rosy~as a multi-stage controller that explicitly encodes a state machine and reacts differently depending on the state\footnote{Classical FRP frameworks tackle this general problem by designing advanced higher-order switching combinators over reactive functions, that are expressively powerful but even less novice-friendly.}, it is useful to lend more structure to the language as the complexity of the controller grows.

Like other FRP languages~\cite{peterson1999lambda}, \rosy~introduces the concept of tasks, as a combination of an initialization step and a continuous controller.
The initializer sets up the stage for the controller, that runs continuously until a predefined terminating event occurs.
For example, we can make the robot turn sideways by a fixed amount of degrees by writing a task:
\begin{example}[Task: Turn left or right]~
\begin{minted}[escapeinside=@@]{haskell}
type Side = Either @\textbf{\color{rosy}{Degrees}}@ @\textbf{\color{rosy}{Degrees}}@

turn :: Side -> @\textbf{\color{rosy}{Task}}@ ()
turn s = @\color{rosy}{task}@ (startTurn s) runTurn

startTurn :: Side -> @\textbf{\color{rosy}{Orientation}}@
          -> @\textbf{\color{rosy}{Memory}}@ @\textbf{\color{rosy}{Orientation}}@
startTurn s o = case s of
  Left a -> @\textbf{\color{rosy}{Memory}}@ (o+@\color{rosy}{degreesToOrientation}@ a)
  Right a -> @\textbf{\color{rosy}{Memory}}@ (o+@\color{rosy}{degreesToOrientation}@ a)

runTurn :: @\textbf{\color{rosy}{Memory}}@ @\textbf{\color{rosy}{Orientation}}@ -> @\textbf{\color{rosy}{Orientation}}@
        -> Either (@\textbf{\color{rosy}{Velocity}}@) (@\textbf{\color{rosy}{Done}}@ ())
runTurn (@\textbf{\color{rosy}{Memory}}@ to) from = if abs d <= err
    then Right (@\textbf{\color{rosy}{Done}}@ ())
    else Left (@\textbf{\color{rosy}{Velocity}}@ 0 (@\color{rosy}{orientation}@ d))
  where d = @\color{rosy}{normOrientation}@ (to-from)
\end{minted}
\end{example}
\noindent
At initialization, the\mintinline{haskell}{ startTurn} reads the robot's \mintinline[escapeinside=@@]{haskell}{@\textbf{\color{rosy}{Orientation}}@} from its odometry information, and writes the desired final \mintinline[escapeinside=@@]{haskell}{@\textbf{\color{rosy}{Orientation}}@} to memory by adding or subtracting the received angle to the current orientation.
The\mintinline{haskell}{ runTurn} controller will rotate the robot towards the desired orientation until the desired and current orientations are equal with a small error margin\mintinline{haskell}{ err}, signalling when it is \mintinline[escapeinside=@@]{haskell}{@\textbf{\color{rosy}{Done}}@}.
The \mintinline[escapeinside=@@]{haskell}{@\textbf{\color{rosy}{Done}}@} type allow returning additional information on task termination. Returning nothing is achieved with the empty type \mintinline{haskell}{()}.

A similar task makes the robot move a fixed distance:
\begin{example}[Task: Move forward or backwards]~
\begin{minted}[escapeinside=@@]{haskell}
data Direction = Forward  @\textbf{\color{rosy}{Centimeters}}@
               | Backward @\textbf{\color{rosy}{Centimeters}}@

move :: Direction -> @\textbf{\color{rosy}{Task}}@ ()
move = ...
\end{minted}
\end{example}
Since tasks can end, in contrast to controllers, we can now mimic the block from Fig.~\ref{fig:block} in \rosy~by using Haskell's monadic notation to sequence tasks in an imperative style:
\begin{example}[Task: Draw a square]~
\begin{minted}[escapeinside=@@]{haskell}
drawSquare :: @\textbf{\color{rosy}{Task}}@ ()
drawSquare = replicateM_ 4 $ do
    move (Forward 2)
    turn (Left 90)
\end{minted}
\end{example}

%% file: env.tex
\section{Environment} \label{sec:ide}

\begin{figure}[t]
    \centering
    \includegraphics[width=\columnwidth]{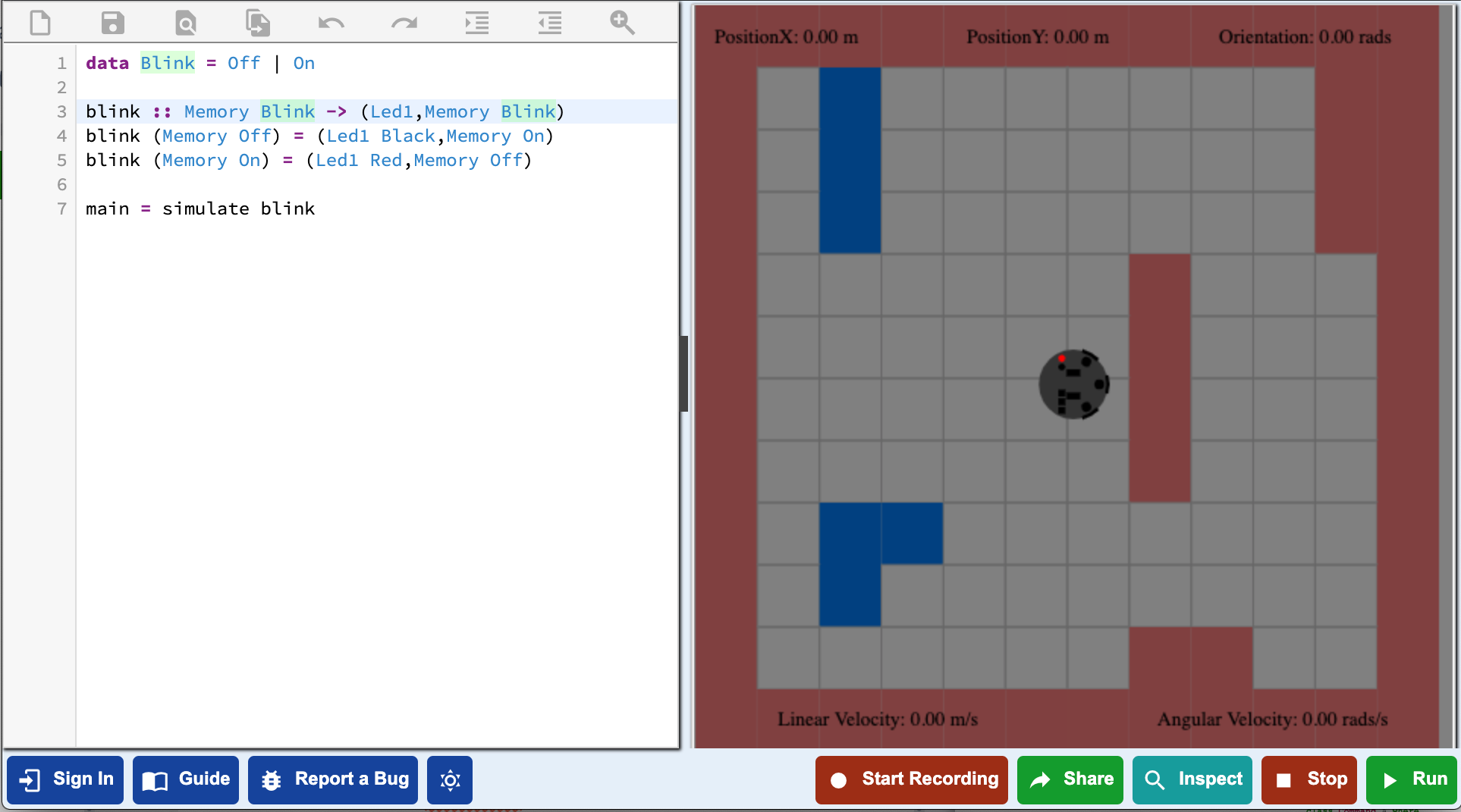}
    \caption{The CodeWorld-powered \rosy~environment.}
    \label{fig:env}
\end{figure}

As well as striving to allow students to learn a ``real'' programming language while freeing them from inessential technical language details outside of how to control a robot, \rosy~comes with a fully integrated development environment.
To really allow students to concentrate on the meaning of their programs, ignoring deployment details, the \rosy~environment runs as a web application inside any modern web browser. 

Figure~\ref{fig:env} shows the \rosy~environment in action.
It is powered by Codeworld~\cite{codeworld}, a modern educational environment for writing graphical Haskell programs such as games and animations. CodeWorld has been used in K12 schools for years, and supports a code editor with features such as syntax highlighting, improved on-the-fly compiler error messages, extensive documentation or easy sharing of projects.

Another vital component of the environment is a visualizer that simulates, directly in the browser, how the controller programmed by the student interacts with a robot in a fictitious 2D world.
At the moment, this is tailored for a TurtleBot2 placed in a tiled world made of floor, walls and cliffs.
In \rosy~training sessions for K12 students that we have hosted at the University of Minho, students could also deploy their same code to control a real robot and perceive the differences between a simulated and a real world.
Enabling students to simultaneously test their examples in simulated and real scenarios played an important role in teaching them the importance of these differences and their influence on the design of approximate, event-driven robotic controllers.

%% file: impl.tex
\section{Under the hood} \label{sec:tech}

\begin{figure}[t]
\includegraphics[width=\columnwidth]{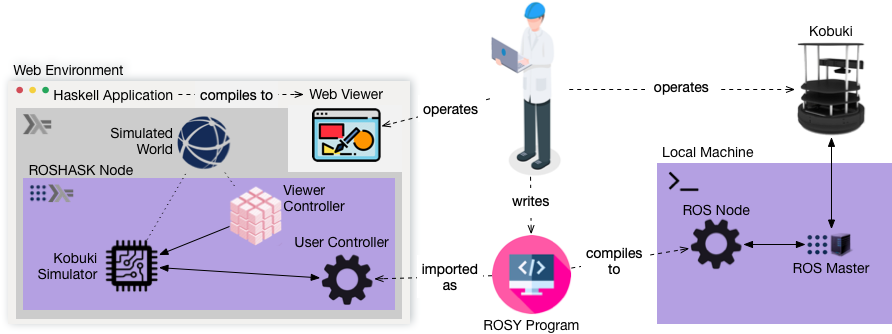}
\caption{The \rosy~architecture.}
\label{fig:arch}
\end{figure}

Despite their simplicity, \rosy~programs are fully compatible with the \ROS~infrastructure.
Under the hood, the \rosy~language is implemented as an embedded domain-specific language that provides an additional abstraction layer over the existing Haskell \ROS~library.
The \rosy~environment is also implemented in Haskell, by extending and specially tailoring the CodeWorld~\cite{codeworld} environment to the \rosy~language. This includes a custom prelude that is imported by default, a custom code pre-processor to automatically derive necessary Haskell type class instances, and a custom graphical simulation of the robot.
All the development source code is open-source and freely available\footnote{\url{https://github.com/hpacheco/codeworld-rosy}}.

\subsection{Haskell \ROS~library}

The \roshask~library~\cite{roshask} enables \ROS~programming within the Haskell ecosystem by supporting the deployment of \ROS~client nodes that are compatible with the \texttt{rostcp} communications protocol.
\roshask~fades the architectural boundaries between \ROS~system and software components, by lifting topics to first-class values and designing a collection of FRP-style combinators to split, fuse and generally manipulate topics in a more expressive, modular and compositional way.

At its core, the library provides functions for subscribing and publishing topics:
\begin{minted}[escapeinside=@@]{haskell}
@\color{ros}{subscribe}@ :: String -> @\textbf{\color{ros}{Node}}@ (@\textbf{\color{ros}{Topic}}@ m a)
@\color{ros}{publish}@   :: String -> @\textbf{\color{ros}{Topic}}@ m a -> @\textbf{\color{ros}{Node}}@ ()
\end{minted}
The type variable\mintinline{haskell}{ m} is a monad for actions with effects, typically \mintinline{haskell}{IO} for interacting with a non-pure outside world, and\mintinline{haskell}{ a} is the type of the subscribed or published event derived from standard \ROS~message type definition files.
Topics are modelled as infinite monadic streams~\cite{perez2018functional}, i.e., monadic actions that produce the next value and a new topic: 
\begin{minted}[escapeinside=@@]{haskell}
newtype @\textbf{\color{ros}{Topic}}@ m a = @\textbf{\color{ros}{Topic}}@ (m (a,@\textbf{\color{ros}{Topic}}@ m a))
\end{minted}
The \mintinline[escapeinside=@@]{haskell}{@\textbf{\color{ros}{Node}}@} monad manages TCP connections and internal buffers of published and subscribed messages.
The following example node subscribes to two sensors, fuses them using the\mintinline[escapeinside=@@]{haskell}{ @{\color{ros}{bothNew}}@} combinator (subsampling the faster topic), maps an action\mintinline{haskell}{ act :: (Sense1,Sense2) -> Cmd} over each pair of sensor values, and publishes the resulting commands:
\begin{minted}[escapeinside=@@]{haskell}
n1 = do t1 <- @{\color{ros}{subscribe}}@ "sense1"
        t2 <- @{\color{ros}{subscribe}}@ "sense2"
        @{\color{ros}{publish}}@ "cmd" $ fmap act $ t1 `@{\color{ros}{bothNew}}@` t2
\end{minted}
Internally, the asynchronous \roshask~behavior is implemented on top of Haskell user-space threads, that are managed by the Haskell runtime and much more efficient than system threads. For instance, a typical\mintinline[escapeinside=@@]{haskell}{ @{\color{ros}{publish}}@} implementation launches a thread that infinitely samples values from a topic and communicates them to the \ROS~master; similarly, merging two topics is done by launching two threads that independently consume each topic and write to a common channel.

Nodes can also be easily composed, for instance, we can simultaneously install a handler that listens to and prints the published commands to the command line:
\begin{minted}[escapeinside=@@]{haskell}
n2 = @{\color{ros}{subscribe}}@ "cmd" >>= @{\color{ros}{runHandler}}@ putStrLn
\end{minted}
In the style of\mintinline[escapeinside=@@]{haskell}{ @{\color{ros}{publish}}@},\mintinline[escapeinside=@@]{haskell}{ @{\color{ros}{runHandler}}@} is a general combinator that launches a thread that will consume values from a topic and execute some user-defined \mintinline{haskell}{IO} action:
\begin{minted}[escapeinside=@@]{haskell}
@{\color{ros}{runHandler}}@ :: (a -> IO b) -> Topic IO a -> @\textbf{\color{ros}{Node}}@ ()
\end{minted}
The composite node\mintinline{haskell}{ n1 >> n2} will communicate the commands produced by\mintinline{haskell}{ n1} both via \texttt{rostcp} and locally to\mintinline{haskell}{ n2}.

\subsection{Architecture}

The \rosy~web environment, depicted on left side of Fig.~\ref{fig:arch}, is designed according to the common Model–View–Controller pattern, with components implemented as separate \roshask~controllers that communicate locally and form a single \roshask~node.
The model controller (loosely) simulates a standard \texttt{kobuki\_node}\footnote{\url{http://wiki.ros.org/kobuki_node}} in a simulated world, the view controller implements a simple 2D animation of the robot within the world, and the user controller represents the node specified by the programmer. To support web-based simulation and loosen the dependency on \ROS, we have adapted the \roshask~library to run without a \ROS~master server. This way, all the Haskell code needed to perform a simulation is compiled via GHCJS into a JavaScript application that  runs directly in the client browser.

Alternatively to web-based simulation, \rosy~ programs can also be executed on a local machine (right side of Fig.~\ref{fig:arch}), by connecting to a \ROS~master server and operate a more realistic Gazebo simulator or a real Kobuki robot. We have tested both scenarios on Ubuntu 14.04 with \ROS~Indigo and a TurtleBot2.

\subsection{Implicit stream programming}

The greatest design decision of the \rosy~language is that controllers have an implicit notion of time.
This favors a simpler declarative style focused on \emph{which} commands are issued \emph{when} events happens, without specifying \emph{how often} subscribers and publishers interact with the \ROS~world, and freeing programmers from common robotic programming details such as clocks, sampling rates or synchronization. Therefore, unlike \roshask, that exposes a full-fledged API to manipulate topics as a whole, \rosy~controllers are less expressive in that they only consider a single point in time.

In \rosy, events are identified by their type. We define two type classes that internally bind \rosy~types and \ROS~message namespaces to streams of subscribed sensors or published commands:
\begin{minted}[escapeinside=@@]{haskell}
class @\textbf{\color{rosy}{Sensor}}@ a where
  @\color{rosy}{sensor}@  :: @\textbf{\color{ros}{Node}}@ (@\textbf{\color{ros}{Topic}}@ IO a)
class @\textbf{\color{rosy}{Command}}@ a where
  @\color{rosy}{command}@ :: @\textbf{\color{ros}{Topic}}@ IO a -> @\textbf{\color{ros}{Node}}@ ()
\end{minted}
For example, the same \mintinline[escapeinside=@@]{haskell}{@\textbf{\color{rosy}{Velocity}}@} type can simultaneously express the act of getting the current velocity from the robot's periodic odometry data and the act of setting the desired velocity by sending a command to the robot's base:
\begin{minted}[escapeinside=@@]{haskell}
instance @\textbf{\color{rosy}{Sensor}}@ @\textbf{\color{rosy}{Velocity}}@ where
  @\color{rosy}{sensor}@ = @{\color{ros}{subscribe}}@ "odom"
    >>= return . fmap (@{\color{ros}{\_twist}}@ . @{\color{ros}{\_twist}}@))
instance @\textbf{\color{rosy}{Command}}@ @\textbf{\color{rosy}{Velocity}}@ where
  @\color{rosy}{command}@ = @{\color{ros}{publish}}@
    "/mobile_base/commands/velocity"
\end{minted}
The \rosy~language currently supports a fixed set of events offered by the \texttt{kobuki\_node} API. Extending support for other robots simply requires defining new boilerplate \mintinline[escapeinside=@@]{haskell}{@\textbf{\color{rosy}{Sensor}}@} and \mintinline[escapeinside=@@]{haskell}{@\textbf{\color{rosy}{Command}}@} instances, as \roshask~already supports many standard \ROS~message types and allows deriving Haskell types from custom \ROS~message files.

To conciliate whole topics with point-wise controllers, we define another type class that implicitly lifts a function on individual values to a controller on streams of values:
\begin{minted}[escapeinside=@@]{haskell}
class @\textbf{\color{rosy}{Controller}}@ a where
  @\color{rosy}{controller}@ :: a -> @\textbf{\color{ros}{Node}}@ ()
\end{minted}
The stream semantics of the lifted controller is then inferred from the function's type signature: inputs correspond to subscribed sensors, and outputs to published commands, e.g.:
\begin{minted}[escapeinside=@@]{haskell}
instance (@\textbf{\color{rosy}{Sensor}}@ a,@\textbf{\color{rosy}{Command}}@ b)
         => @\textbf{\color{rosy}{Controller}}@ (a -> b) where
  @\color{rosy}{controller}@ f = @\color{rosy}{sensor}@ >>= @\color{rosy}{command}@ . fmap f
\end{minted}
Instances for composite types perform implicit stream programming. For example, a \mintinline[escapeinside=@@]{haskell}{@\textbf{\color{rosy}{Controller}}@} that receives an input pair is fusing data from two sensors:
\begin{minted}[escapeinside=@@]{haskell}
instance (@\textbf{\color{rosy}{Sensor}}@ a,@\textbf{\color{rosy}{Sensor}}@ b) => @\textbf{\color{rosy}{Sensor}}@ (a,b) where
  @\color{rosy}{sensor}@ = liftM2 @{\color{ros}{bothNew}}@ @\color{rosy}{sensor}@ @\color{rosy}{sensor}@
\end{minted}
As another example, a \mintinline[escapeinside=@@]{haskell}{@\textbf{\color{rosy}{Controller}}@} that returns possibly different commands is splitting the output stream (using the\mintinline[escapeinside=@@]{haskell}{ @{\color{ros}{tee}}@} \roshask~combinator that duplicates a topic), and processing each type of commands independently:
\begin{minted}[escapeinside=@@]{haskell}
instance (@\textbf{\color{rosy}{Command}}@ a,@\textbf{\color{rosy}{Command}}@ b)
         => @\textbf{\color{rosy}{Command}}@ (Either a b) where
  @\color{rosy}{command}@ t = do (t1,t2) <- @{\color{ros}{tee}}@ t
                 @\color{rosy}{command}@ $ lefts t1
                 @\color{rosy}{command}@ $ rights t2
\end{minted}
Multiple \mintinline[escapeinside=@@]{haskell}{@\textbf{\color{rosy}{Controller}}@}s are executed in parallel threads, and can be composed in sequence:
\begin{minted}[escapeinside=@@]{haskell}
instance (@\textbf{\color{rosy}{Controller}}@ a,@\textbf{\color{rosy}{Controller}}@ b)
         => @\textbf{\color{rosy}{Controller}}@ (a,b) where
  @\color{rosy}{controller}@ (a,b) = @\color{rosy}{controller}@ a >> @\color{rosy}{controller}@ b
\end{minted}

\subsection{Supporting user-defined events and memory}

\rosy~also allows the declaration of user-defined data types for more modular intra-node communication.
Since these need not be bound to \ROS~namespaces, every newly declared user-defined data type is by default a \mintinline[escapeinside=@@]{haskell}{@\textbf{\color{rosy}{Sensor}}@} and a \mintinline[escapeinside=@@]{haskell}{@\textbf{\color{rosy}{Command}}@}.
This is supported by a new \mintinline[escapeinside=@@]{haskell}{@\textbf{\color{rosy}{UserNode}}@} monad that extends \mintinline[escapeinside=@@]{haskell}{@\textbf{\color{ros}{Node}}@} capabilities with local type-indexed event buffers for user-defined data types.

Since \rosy~programs are by design pure Haskell functions, there is no native support for common robotics design patterns that global memory. It is nonetheless possible to emulate global variables by publishing an initial event, and on every update subscribing to current event and re-publishing its updated value. Even so, this pattern can be error-prone as the programmer needs to be cautious about subscribing and publishing the ``variable'' the right number of times in order to keep it alive. This may also be problematic if more than one controller is manipulating the same ``variable'' in parallel.

To avoid these caveats, \rosy~supports node-specific memory, distinguishable through a type-level tag:
\begin{minted}[escapeinside=@@]{haskell}
data @\textbf{\color{rosy}{Memory}}@ a = @\textbf{\color{rosy}{Memory}}@ a
\end{minted}
We implement memory by extending the \mintinline[escapeinside=@@]{haskell}{@\textbf{\color{rosy}{UserNode}}@} monad with a transactional memory store, holding a global value for every distinct type\mintinline{haskell}{ a}.

In order to support transactional controllers, or more specifically, be able to execute each controller thread as a single transaction, we must generalize our sensor and command interfaces to produce and consume topics of transactions\footnote{The \mintinline{haskell}{STM} monad stands for Haskell's software transactional memory library. The \mintinline{haskell}{Maybe} type is a technical requirement for filtering values inside a transaction, since instances must not change the periodicity of the topics.}:
\begin{minted}[escapeinside=@@]{haskell}
@\color{rosy}{sensor}@   :: @\textbf{\color{rosy}{UserNode}}@ (@\textbf{\color{ros}{Topic}}@ IO (STM a))
@\color{rosy}{command}@  :: @\textbf{\color{ros}{Topic}}@ IO (STM (Maybe a))
         -> @\textbf{\color{rosy}{UserNode}}@ (@\textbf{\color{ros}{Topic}}@ IO (STM ()))
\end{minted}
A \mintinline[escapeinside=@@]{haskell}{@\textbf{\color{rosy}{Sensor}}@ (@\textbf{\color{rosy}{Memory}}@ a)} returns a topic that  repeatedly reads from the memory variable of type\mintinline{haskell}{ a}, while a \mintinline[escapeinside=@@]{haskell}{@\textbf{\color{rosy}{Command}}@ (@\textbf{\color{rosy}{Memory}}@ a)} appends memory writes to a topic of transactions, returning a new topic.
The greatest change occurs on commands that, unlike before, must be published synchronously within the same transaction, meaning we can no longer fork a topic and publish each side independently.
We can execute a controller thread by installing a handler that\mintinline{haskell}{ atomically} executes each transaction as a side effect:
\begin{minted}[escapeinside=@@]{haskell}
instance (...) => @\textbf{\color{rosy}{Controller}}@ (a -> b) where
  @\color{rosy}{controller}@ f = @\color{rosy}{sensor}@ >>= @\color{rosy}{command}@ . fmap f
               >>= lift . @{\color{ros}{runHandler}}@ atomically
\end{minted}

\subsection{Sequencing tasks}

In \roshask~and other FRP approaches, topics are modelled as  infinite streams and topic handlers are program-long threads continuously waiting on and reacting to events.
The fact that the data flow graph, inferred from the wiring of stream combinators, is typically known statically, allows \roshask~to register all subscribers and publishers with the \ROS~master at node initialization, before starting to actually process data.

In \rosy, a task is defined as an initialization action, a continuous controller, and a terminating event:
\begin{minted}[escapeinside=@@]{haskell}
@\color{rosy}{task}@ :: (@\textbf{\color{rosy}{Command}}@ init,@\textbf{\color{rosy}{Controller}}@ ctrl)
     => init -> ctrl -> @\textbf{\color{rosy}{Task}}@ end
\end{minted}
A controller issues termination via a special event type:
\begin{minted}[escapeinside=@@]{haskell}
data @\textbf{\color{rosy}{Done}}@ a = @\textbf{\color{rosy}{Done}}@ a
\end{minted}
We also make \mintinline[escapeinside=@@]{haskell}{@\textbf{\color{rosy}{Task}}@} a monad, so that programmers can use monadic notation to sequence tasks. For composing two smaller tasks in into a composite task, where the output of the first is passed on to the second, we may write:
\begin{minted}[escapeinside=@@]{haskell}
task12 = do end1 <- @\color{rosy}{task}@ init1 ctrl1
            @\color{rosy}{task}@ (init2 end) ctrl2
\end{minted}
In this scenario, controllers no longer run forever: when the first task ends, we must uninstall the controller\mintinline{haskell}{ ctrl1} and install a new controller\mintinline{haskell}{ ctrl2} for the second task.
We have extended \roshask~to support dynamic node configuration:
each task runs within its own \mintinline[escapeinside=@@]{haskell}{@\textbf{\color{rosy}{UserNode}}@}; publishers and subscribers are registered at declaration time; tasks keep a fine-grained control of launched threads, and all children threads are killed when exiting the parent \mintinline[escapeinside=@@]{haskell}{@\textbf{\color{rosy}{UserNode}}@}\footnote{Note that messages are not lost when transitioning between tasks, since a \mintinline{haskell}{Node} keeps global buffers of published and subscribed \ROS~topics.}.